\documentclass[aps,pre,reprint,groupedaddress,longbibliography,nofootinbib]{revtex4-1}
\usepackage{times}
\usepackage{chemarrow} % for xarrows 
\usepackage{amsmath,amssymb,mathrsfs}
\usepackage{mathtools}
\usepackage{graphicx,epsfig}
\usepackage{xcolor}
\usepackage{bbm}
\usepackage{extarrows,chemarrow,xypic}
\usepackage{chemarrow}
\usepackage{hyperref}
\usepackage{microtype}
\DisableLigatures[f]{encoding = *, family = *}
\usepackage[none]{hyphenat}
\usepackage{lipsum}
\def\rd{{\rm d}}

\def\vx{{\bf x}}

\def\mQ{{\bf Q}}

\def\mPi{\mbox{\boldmath$\Pi$}}

\begin{document}

%\preprint{}

\title{Duality Symmetry, Two Entropy Functions, and an Eigenvalue Problem in Gibbs' Theory}



\author{Jeffrey Commons}
\affiliation{Department of Physics, University of Washington, Seattle, WA 98195-1560, USA}

\author{Ying-Jen Yang}
\affiliation{Department of Applied Mathematics, University of Washington,
Seattle, WA 98195-3925, USA}

\author{Hong Qian}
\affiliation{Department of Applied Mathematics, University of Washington,
Seattle, WA 98195-3925, USA}
%\email[]{}
%\thanks{}


\date{\today}

\begin{abstract}
We generalize the convex duality symmetry in Gibbs' statistical ensemble formulation, between Massieu's free entropy $\Phi_{V,N}(\beta)$ and the Gibbs entropy $\varphi_{V,N}(u)$ as a function of mean internal energy $u$.  The duality tells us that Gibbs thermodynamic entropy is to the law of large numbers (LLN) for arithmetic sample means what Shannon's information entropy is to the LLN for empirical counting frequencies.  Following the same logic, we identify $u$ as the conjugate variable to counting frequency, a Hamilton-Jacobi equation for Shannon entropy as an equation of state, and suggest an eigenvalue problem for modeling statistical frequencies of correlated data.
\end{abstract}

%\keywords{}

\maketitle


%\section{}

%\tableofcontents

	In recent years, through activities in stochastic thermodynamics and applying the mathematical theory of large deviations to the asymptotic behavior of the LLN, ``the essence of a thermodynamic description is not found in its connection to conservation laws, microscopic reversibility, or the equilibrium state relation they entail, despite the central role those play'' in the current teaching; see \cite{esmith} and the references cited within.  In addition to the derivation of fluctuation relations in connection to various nonequilibrium entropy productions, more recent results on the unification of stochastic chemical kinetics and Gibbsian thermochemistry \cite{ge-qian-16}, as well as the probabilistic elucidation of T. L. Hill's thermodynamics of small systems (1963) \cite{hill-book,lu-qian-20}, have all further substantiated the above claim.

The Large Deviations Theory (LDT) from probability \cite{oono,dembo} provides the mathematical underpinning for understanding statistical {\em observables}, of which the arithmetic mean values  and empirical relative frequencies (proportions) are the two most important types. They have asymptotic behaviors that are independent of the details of a specific probabilistic model \cite{vulpiani-book}; in fact we shall show that the Gibbsian statistical thermodynamics can be framed, independent of mechanics, as the asymptotic statistical laws for counting frequencies. The mathematics reveals an important duality structure between two functions, a large deviation rate function (LDRF) and its cumulant generating function (CGF).  In statistical mechanics, they correspond to Gibbs' entropy as a function of all extensive variables and the Massieu-Guggenheim entropy as a function of all intensive, conjugate variables \cite{lu-qian-20}.  

LDT offers a clear contradistinction between the two best-known entropies, that of Gibbs and that of Shannon \cite{shannon-book}.  Although writings on the subject are already vast,  here we give a more coherent presentation showing that the Gibbs entropy, defined via temperature derivative of canonical partition function as in standard textbooks, is to the LLN for arithmetic mean values what the Shannon entropy is to the LLN for empirical counting frequencies.  They are known as Cr\'{a}mer's theorem and the Sanov theorem in mathematics respectively \cite{dembo}.  

	With this newfound perspective and by explicating the duality for the Shannon-Sanov relative entropy \cite{hobson,shore-johnson}, we show that the conjugate variables to ``relative  frequencies'' $\{\nu_i\}$ are a set of ``energy parameters'' in $k_BT$ units, $\{u_j\}$.  The duality turns out to be precisely Gibbs' canonical theory. We further consider repeated measurements on a time-correlated Markov process and find a novel eigenvalue problem for $\{\sqrt{\nu_i}\}$.

{\bf\em Duality symmetry between $\psi_{V,N}(\beta)$ and $\varphi_{V,N}(u)$ |} It is important to recall that as thermodynamic potential functions, the Gibbs entropy is a function of an extensive variable,  the internal energy, and Helmholtz free energy or Massieu free entropy are functions of the intensive conjugate variable $T$.  There is a convex duality between these two functions \cite{rockafellar}. 

	Consider a mechanical system consisting of $N$ particles, $\vx=(x_1,\cdots,x_N)$, in a container with volume $V$ under temperature $T=(k_B\beta)^{-1}$, with a potential energy function $U_N(\vx)$. The logarithm of the canonical partition function
\begin{equation}
           \Phi_{V,N}(\beta) = k_B\log \int_{V^N} e^{-\beta U_N(\vx)}
              \rd\vx
\end{equation}
has been known as Massieu's free entropy.  See \cite{lu-qian-20} for a very recent discussion.  Then the equilibrium entropy of the statistical mechanical system according to Gibbs' theory is
\begin{eqnarray}
   S(\beta) &=& \frac{\partial [\beta^{-1}\Phi_{V,N}(\beta)]}{\partial\beta^{-1}} 
= -\beta\frac{\partial\Phi_{V,N}}{\partial\beta}+\Phi_{V,N}(\beta) ,
\end{eqnarray}
which can be expressed in terms of Legendre-Fenchel transform (LFT) from convex analysis:
\begin{subequations}
\label{equation3}
\begin{eqnarray}
	\varphi_{V,N}(u) &\equiv& -S\big(\beta(u)\big) = 
     -\inf_{\beta}\Big\{ \beta u+\Phi_{V,N}(\beta)  \Big\} 
\\
	&=& \left\{
  \begin{array}{l} 
      \displaystyle
    -\beta u -\Phi_{V,N}(\beta)   \\[6pt]
      \displaystyle
    u(\beta) = -\partial\Phi_{V,N}(\beta)/\partial\beta,
       \end{array} \right.
\end{eqnarray}
\end{subequations}
in which function $u(\beta)$ is monotonic, therefore 
$\beta(u)$ is well defined.  Since $\Phi_{V,N}(\beta)$ is a convex function of $\beta$, Eq. \ref{equation3} has a dual expression:
\begin{equation}
     \Phi_{V,N}(\beta) = -\inf_{u} \Big\{ \beta u+
          \varphi_{V,N}(u)   \Big\},
\label{eq4S}
\end{equation} 
and a corresponding pair of differential equations showing $\varphi$ and $\Phi$ as thermodynamic potential functions for their respective independent variables, with corresponding ``forces'' $\beta$ and $u$:
\begin{subequations}
\label{dualrel}
\begin{eqnarray}
  \rd\varphi_{V,N}(u)  &=& \left(\frac{\rd\varphi_{V,N}(u)}{\rd u}\right) \rd u = -\beta(u) \rd u,
\\
	\rd\Phi_{V,N}(\beta) &=& \left(\frac{\rd\Phi_{V,N}(\beta)}{\rd \beta}\right) \rd\beta =  -u(\beta) \rd\beta.
\end{eqnarray}
\end{subequations}
The duality in Eq. \ref{dualrel} can be further extended to intensive conjugate variables $p$ and $\mu$ which correspond to extensive variables $V$ and $N$.  This leads to a duality symmetry between the fundamental thermodynamic relation and the Hill-Gibbs-Duhem (HGD) equation in nanothermodynamics \cite{lu-qian-20}, respectively.  In the classical macroscopic thermodynamic limit, the HGD equation becomes the Gibbs-Duhem equation, with the duality symmetry broken: When $\varphi_{V,N}(u)$ is an Eulerian homogeneous degree-$1$ function of $u,V,N$, its complete LFT $\psi(\beta,p,\mu)$ becomes singular except on a hyper-surface that yields the equation of state (EoS):
\begin{equation}
                  \mu= \frac{1}{N}\sup_{V}\Big\{ pV - 
              \beta^{-1}\Phi_{V,N}(\beta) \Big\}.
\label{eq4mu}
\end{equation}
The $\mu$ is a function of $p$ and $\beta$; 
the $\psi(\beta,p,\mu)$ vanishes. In the familiar standard thermodynamics notations, ${\Phi=-F/T=-(U-TS)/T}$ where ${1/T=\partial S/\partial U}$, ${\mu=G/N=(pV+F)/N}$ where $p=-\partial F/\partial V$. They correspond to Eqs. \ref{eq4S} and \ref{eq4mu}, with the infimum and supremum carried out by differentiating convex functions.

{\bf\em The Gibbs and the Shannon-Sanov entropies, currently their relation |} For a given probability distribution over a set of $n$ discrete events with probability mass function $p_i$, $1\le i\le n$, Shannon's entropy for empirical frequencies $\{\nu_i\}$ is a special case of the more general relative entropy \cite{hobson,shore-johnson}
\begin{equation}
\label{eq7}
        H\big(\{\nu_i\}\big)=\sum_{i=1}^n \nu_i\log\frac{\nu_i}{p_i} \ \text{ or }
         \int_{\mathbb{R}} f(x)\log\frac{f(x)}{p(x)}\rd x.
\end{equation}
Shannon's original formula is taken relative to a uniform probability:
$-H(\nu)=-\sum_{i=1}^n\nu_i\log\nu_i -\log n$.  There is an extra, non-consequential constant $-\log n$.  The first term is interpreted as the ``missing information'' w.r.t. the maximum information $-\log n$.

According to standard textbooks on equilibrium statistical mechanics \cite{kersonhuang}, the Gibbs entropy is defined by starting from a given density of states in terms of energy $\Omega(u)$, $u\in\mathbb{R}$, 
\begin{equation}
	F(\beta) = -\beta^{-1} \log\int_{\mathbb{R}} \Omega(u)
         e^{-\beta u}\rd u,\
	S = -\frac{ \partial F(\beta) }{\partial (1/\beta)},
\end{equation}
where $S$ is the Gibbs entropy. $F$ is Helmholtz free energy and $\beta=(k_BT)^{-1}$ is inverse temperature.  A {\em canonical distribution} is
\[
		f(E) = \Omega(u) e^{-\beta E}\left(\int_{\mathbb{R}} \Omega(E)
                e^{-\beta E}\rd E\right)^{-1},
\]
from which one has the well-known equation
\begin{equation}
   S = -\frac{ \partial F(\beta) }{\partial (1/\beta)} 
    =  -\int_{\mathbb{R}} f(E)\log\left(\frac{f(E)}{\Omega(E)}
          \right)\rd E.
\end{equation}
If one identifies $\Omega(E)$ with the uniform density $1/N$ and integration w.r.t. $E$ with summation over $n$; then the Gibbs entropy and the Shannon entropy are considered to be the {\em same}.
%\\
%	&=&  \log\int_{\mathbb{R}} \Omega(E)
%        e^{-\beta E}\rd E - \beta \frac{\partial}{\partial\beta}
%             \log\int_{\mathbb{R}} \Omega(E)
%         e^{-\beta E}\rd E
%\\
%	&=&  \log\int_{\mathbb{R}} \Omega(E)
%         e^{-\beta E}\rd E +\beta \frac{\displaystyle 
%               \int_{\mathbb{R}}  E \Omega(E)
%         e^{-\beta E}\rd E}{
%     \displaystyle \int_{\mathbb{R}} \Omega(E)
%         e^{-\beta E}\rd E }
%\\
%	&=&  \log\int_{\mathbb{R}} \Omega(E)
%         e^{-\beta E}\rd E - \frac{\displaystyle 
%               \int_{\mathbb{R}}  \log e^{-\beta E} \Omega(E)
%         e^{-\beta E}\rd E}{
%     \displaystyle \int_{\mathbb{R}} \Omega(E)
%         e^{-\beta E}\rd E }
%\\
%	&=& - \frac{\displaystyle 
%               \int_{\mathbb{R}}  \log \frac{ e^{-\beta E}  {\int_{\mathbb{R}} \Omega(E)
%         e^{-\beta E}\rd E }\ \Omega(E)
%         e^{-\beta E}\rd E}{
%     \displaystyle \int_{\mathbb{R}} \Omega(E)
%         e^{-\beta E}\rd E }

{\bf\em Beyond the current understanding |}
One has found an agreement between the mathematical expressions for the two entropies.  However, a fundamental question remains unanswered: Why should the mathematical function in (\ref{eq7}) be minimized? Though from thermodynamics originally, this {\em principle} must have a deeper root in statistics; the answer has to be found in the mathematical logic of probability. 

There is a new principle that calls for a revisit of the two entropies: {\em The entropy function arises as the asymptotic exponent for a law of large numbers}.  There are many different laws of large numbers, thus there are many different forms of entropy function, c.f. Eq. \ref{I4ctMC} below.  Gibbs' and Shannon's are merely two special ones for the two most widely studied LLN.  We shall consider two i.i.d. sampling problems: (i) A discrete random event with probability mass function $\pi_k$, $1\le k\le n$, and (ii) a continuous real-valued random variable $X$ with probability density function $\Omega(x)$.

{\bf\em Shannon's relative entropy as a LDRF |}
For the discrete problem, let us assume that within total $M$ i.i.d. samples, there are $m_k$ number of observations of the $k^{th}$ event: ${m_1+\cdots + m_n = }M$.  It is clear that if one carries out another $M$ observations, one would obtain another set of ${\{m_k;1\le k\le n\}}$.  So the $m_k$'s are themselves random variables with a
joint probability distribution
\begin{equation}
       \frac{M!}{m_1!\cdots m_n!}\pi_1^{m_1}\cdots\pi_n^{m_n}.
\label{h1}
\end{equation}
As $M\to\infty$, the relative frequency $\nu_k:=\frac{m_k}{M}\to\pi_k$: The proportion of observing the $k^{th}$ event approaches to its probability.  This is known as Borel's LLN.

	The {\em Large Deviation Principle} says that the distribution in (\ref{h1}), as $M\to\infty$, has an asymptotic expression
$e^{-M\varphi(\nu_1,\cdots,\nu_n)}$, where $\varphi(\nu_1,\cdots,\nu_n)$ is exactly the $H(\{\nu_i\})$ in Eq. \ref{eq7}, with $p_k=\pi_k$.

{\bf\em The Gibbs entropy is also a LDRF |}
Now for the continuous problem, let $X_1,X_2,\cdots, X_M$ be $M$ i.i.d. samples of a random variable $X$, with probability density function $\Omega(x)$.  Then the LLN for the {\em mean value} says that
\[
    \overline{X}_M := 
   \frac{X_1 + \cdots + X_M}{M} \to \mathbb{E}[X] =
  \int_{\mathbb{R}} x\Omega(x)\rd x,
\]
as $M\to\infty$.  It is clear that if one obtains another $M$ samples, the $\overline{X}_M$ will be different.  So $\overline{X}_M$ itself is a 
random variable with a probability density function
$f_{\overline{X}_M}(x)$.

	Again, the LDT says that as $M\to\infty$,
$f_{\overline{X}_M}(x)$ has an asymptotic expression $e^{M\psi(x)}$.  To compute the exponent $\psi(x)$, we employ the method of CGF,
\begin{equation}
     \Lambda(\beta) = \log\int_{\mathbb{R}} \Omega(x)e^{-\beta x}\rd x,
\end{equation}
and note that the CGF of $\overline{X}_M$ is
$\Lambda_{\overline{X}_M}(\beta)=M\Lambda(\beta/M)$. Then
\begin{equation}
    \Lambda(\beta)=  
 \frac{1}{M}\Lambda_{\overline{X}_M}\big(M\beta\big)  
                           = \sup_x\big\{\psi(x) -\beta x \big\},
 \end{equation}
which implies
\begin{equation}
    \psi(x) = \inf_{\beta} \big\{ \beta x+\Lambda(\beta)\big\}.
\label{g4}
\end{equation}
Now to carry out the optimization problem in (\ref{g4}), let us use calculus to find the optimal $\beta^*$: $x+\Lambda'(\beta)=0$.  Therefore,
\begin{equation}
 \left\{\begin{array}{ccl}
	\psi &=&-\beta \Lambda'(\beta) + \Lambda(\beta)
	= \displaystyle \frac{\partial}{\partial(1/\beta)}\Big[\beta^{-1}\Lambda(\beta) \Big],
\\
	x &=&  \displaystyle -\frac{\partial}{\partial\beta}\Lambda(\beta).
    \end{array} \right. 
\label{g5}
\end{equation}
Eq. (\ref{g5}) gives the function $\psi(x)$ in a parametric form.
We see that if we identify $-\beta^{-1}\Lambda(\beta)$
as $F(\beta)$, then the $\psi$ in (\ref{g5}) is the Gibbs entropy, and $x$ is the internal energy.

	Therefore, as the asymptotic exponent for the probability of large deviations, {\em the Shannon entropy is to Borel's LLN for empirical counting frequencies what the Gibbs entropy is to the standard LLN for arithmetic mean values.}

{\bf\em Gibbs-Kirkwood potential energy and the dual to the Shannon-Sanov entropy |} Through Legendre transforms, the theory of classical thermodynamics introduces intensive conjugate variables to extensive quantities, such as $\beta$ to internal energy $U$, $p$ to volume $V$, and $\mu$ to particle number $N$.  The same conjugate variables enter Gibbs' theory of statistical mechanics via Laplace transforms of various partition functions corresponding to extensive variables.  The result from the previous sections shows clearly that a duality structure between the LDRF and its CGF plays a central role in a new understanding of Gibbs' and Shannon's entropies.

By the same logic the following question immediately arises: What is the corresponding CGF for Shannon's relative entropy, i.e., Sanov's LDRF in (\ref{eq7})? We carry out its LFT using $u_i$ to denote the conjugate variable of $\nu_i$ under the normalization constraint $\sum_i^n\nu_i=1$:
\begin{eqnarray}
   \Xi\big(\{u_i\}\big) &=& \inf_{\{\nu_i\}}
    \left\{ \sum_{i=1}^n u_i\nu_i + H\big(\{\nu_i\}\big) \right\}
\label{gkpe}\\
	  &=& \sum_{i=1}^n \nu_i^* \log\left(\frac{\nu_i^*}{p_i e^{-u_i} }\right)
	   =  \lambda = -\log\sum_{i=1}^n p_ie^{-u_i},
\nonumber
\end{eqnarray}
in which $\lambda$ appears as the Lagrange multiplier for the constrained optimization
${\nu_i^*=p_ie^{-u_i+\lambda}}$.
For fixed $u_i$, the $\nu_i$ are determined from ${u_i = -\partial H(\{\nu_i\})/\partial \nu_i+\lambda}$. The additive constant $\lambda$ to the $u_i$'s represents a gauge freedom, a feature that is absent in the traditional Legendre transform.  Eq. \ref{gkpe} has the exact form of Helmholtz free energy from standard statistical mechanics, when a set of ``energy'' parameters in $k_BT$ units, $\{u_i\}$ is given:
\begin{equation}
	 \Xi\big(\{u_i\}\big)  = \underbrace{
         \sum_{i=1}^n \nu_i^*u_i  }_{ \text{mean internal energy} } - \underbrace{ \left(-\sum_{i=1}^n\nu^*_i\log\frac{\nu^*_i}{p_i} \right) }_{\text{entropy}}.
\label{17}
\end{equation}
More precisely, the mean internal energy term in (\ref{17}) should be understood as the mean value of Kirkwood's potential of mean force $u_i$ \cite{kirkwood}.  The conjugate variables $u_i$ therefore are interpreted as a ``potential function'' that represents the statistical deviations of the empirical frequencies $\{\nu_i\}$ away from the true probability $\{p_i\}$ in the large-deviation scenario. 

We shall call the set of $\{u_i\}$ introduced in (\ref{gkpe}), the conjugate variables to the counting frequencies $\{\nu_i\}$, {\em Gibbs-Kirkwood potential energies}; they comes with a ``natural'' additive constant $\lambda$ due to the normalization of frequencies $\{\nu_i\}$.  One then recovers the duality:
\begin{subequations}
\begin{eqnarray}
   \rd H\big(\{\nu_i\}\big) &=& \sum_{i=1}^n \left(
 \frac{\partial H}{\partial\nu_i}\right)\rd \nu_i = 
      - \sum_{i=1}^n u_i \big(\{\nu_i\}\big) \rd\nu_i,
      \hspace{0.7cm}
\\
    \rd\Xi\big(\{u_i\}\big) &=& \sum_{i=1}^n
          \left(\frac{\partial\Xi}{\partial u_i}\right) \rd u_i
          = \sum_{i=1}^n
             \nu_i\big(\{u_i\}\big)\rd u_i,
\end{eqnarray}
\end{subequations}
where $u_i(\{\nu_i\})=$ $-\log(\nu_i/p_i)+\lambda$.
  
The Massieu free entropy as a function of $\beta$ is the LFT of the Gibbs entropy as a function of energy; the Helmholtz free energy as a function of $\{u_i\}$ is the LFT of Shannon-Sanov entropy as a function of frequencies.

{\bf\em Entropy analysis as in thermodynamics |}
To compare and contrast with statistical thermodynamics, we consider a proper entropy function of non-normalized absolute frequencies $\{m_i\}$, $S(\{m_i\})=-MH(\{m_i/M\})$, which is an Eulerian homogeneous function of degree $1$: ${S\big(\{\alpha m_i\}\big)=\alpha S(\{m_i\})}$.  
Its Legendre transform yields a singular function:
\begin{equation}
       u_n = 
       \frac{1}{m_n}\underbrace{ \left( S\big(\{m_i\}\big) - \sum_{j=1}^{n-1} u_jm_j \right) }_{\text{ a function of
       $u_1,\cdots,u_{n-1}$, and $m_n$}},
\end{equation}
which resembles the Gibbs-Duhem equation, with both integral and differential forms:
\begin{subequations}
\label{22}
\begin{eqnarray}
      &&  S\big(\{m_i\}\big) = \sum_{i=1}^n m_i u_i,
\\
      &&  m_i\big(\{u_k\}\big) = 
          \frac { p_i e^{-u_i} } { Z } \text{ where }
 Z = \frac{\sum_{i=1}^n p_ie^{-u_i} }{  \sum_{j=1}^n m_j }, \hspace{1cm}
\\
	  && \sum_{i=1}^n m_i\big(\{u_k\}\big)\rd u_i
           =  \sum_{i=1}^n \rd m_i\big(\{u_k\}\big) = 0.
\end{eqnarray}
\end{subequations}
We have just generalized the concepts of Gibbs function and chemical potentials in thermochemistry to statistical absolute counting frequencies, in terms of an entropy function $S(\{m_i\})$ and the Gibbs-Kirkwood potential functions as the conjugate variables $\{u_i\}$.  Fixing the gauge $\lambda = 0$ gives the EoS, a stationary Hamilton-Jacobi equation of the entropy function $S$:
\begin{equation}
    \sum_{i=1}^n p_ie^{-u_i} - 1 = 0, \
            u_i = \frac{\partial S(\{m_i\})}{\partial m_i}.
\end{equation}

{\bf\em Markov chain large deviations and conjugate process |} Beyond the i.i.d. empirical counting, most repeated measurements of a stationary stochastic signal have a temporal correlation which can be represented by a Markov process $X_t$, $t=0,1,\cdots$, with transition probability matrix $\{p_{ij}\}$ between consecutive measurements.  The LDRF of the counting  frequencies along a path is \cite{dembo}:
\begin{equation}
  I(\{\nu_i\})
       = - \inf_{\text{all }\mu_i>0} \sum_{i=1}^n \nu_i\log\left(
           \frac{1}{ \mu_i }\sum_{j=1}^n \mu_j p_{ji} \right).
\label{I4ctMC}
\end{equation}
The infimum in (\ref{I4ctMC}) can be solved from a system of nonlinear equations for the $\mu_i$'s:
\begin{subequations}
\label{equation25}
\begin{eqnarray}
  \sum_{i=1}^n \nu_i \left( \frac{\mu_kp_{ki}}{\sum_{j=1}^n \mu_jp_{ji} }
              \right) = \nu_k, \
      1\le k\le n.
\\
    I\big(\{\nu_i\}) = \sum_{i=1}^n \nu_i
          \log\left(\frac{\nu_i}{v_i}\right), \ 
           \nu_i = v_i\sum_{j=1}^n 
             \frac{ \mu_jp_{ji} }{\mu_i}. 
\end{eqnarray}      
\end{subequations}
Eq. \ref{equation25} has two very different interpretations: In data statistical terms, the observed relative frequency $\{\nu_i\}$ is the stationary distribution of a Markov chain with transition probability $\hat{p}_{ij}$:
\begin{equation}
        \hat{p}_{ij}= \frac{\mu_jp_{ji} }{\sum_{k=1}^n\mu_k p_{ki}}  
                        =\frac{ \mu_j\Pr\{X_{t+1}=i|X_t=j\} }{
                 \Pr\{ X_{t+1}=i\} },
\end{equation}
which is the Bayesian posterior distribution for $X_t$ given $X_{t+1}=i$, under the original Markov process with ${\Pr\{X_t=j,X_{t+1}=i\}=\mu_jp_{ji}}$. (\ref{equation25}a) finds the hidden $\{\mu_i\}$ from the observed $\{\nu_i\}$.  Eq. \ref{equation25}b also has a thermodynamic interpretation: $\log(\nu_i/v_i)$ in (\ref{equation25}b) is a Gibbs-Kirkwood potential representing the deviation of observed statistical frequency $\{\nu_i\}$ from the stationary probability $\{\pi_i\}$.  It is zero when $\nu_i=\pi_i$ $\forall i$, which yields $\mu_k=\pi_k$ and $\nu_i=v_i$ from (\ref{equation25}). In terms of $\nu$'s and $v$'s,
$\hat{p}_{ij}=$ $(\mu_jp_{ji}/\mu_i) v_i/\nu_i$.

From a mathematical modeling standpoint, finding $\{p_{jk}\}$ for an underlying Markov system and finding a conjugate process $\{\hat{p}_{jk}\}$ for the {\em empirical frequency of measurements} on the stochastic system are equivalent. Any given set of observed relative frequencies $\{\nu_i\}$ is {\em represented} as the stationary distribution of the conjugate Markov process w.r.t. a hidden measure $\{\mu_i\}$ that serves as a tilting \cite{barato}, and represents the {\em measurement} on probability.

	For a continuous-time Markov chain with $p_{ij}=(e^{\mQ\Delta t})_{ij}$
where $\mQ=\{q_{ij}\}$ is a transition rate matrix, The conjugate process w.r.t $\{\mu_i\}$ has transition probability rate matrix
\begin{equation}
      \hat{q}_{ij} = \frac{\mu_jq_{ji}}{\mu_i} \text{ if }  i\neq j, \ \hat{q}_{ii} =
     -\sum_{j\neq i}^n \hat{q}_{ij},
\label{equation34}
\end{equation}
and $\{\nu_i\}$ is its stationary measure.  Actually ${\hat{q}_{ij}=\mu_jq_{ji}/\mu_i-r_i\delta_{ij}}$ where $r_i=\nu_i/v_i$ has another interpretation: the {\em instantaneous rate of change} in the  logarithmic probability density of state $i$ following the Markov process: $\rd\log\mu_i/\rd t = \sum_{k=1}^n \mu_kq_{ki}/\mu_i=r_i$.
When $\forall i$ $\mu_i=\pi_i$, $r_i=0$.  The larger a $|r_i|$ value for a state $i$, the faster the state probability relaxes to $\pi_i$.  The Gibbs-Kirkwood potential for a Markov chain also acquires a kinetic meaning.

{\bf\em Markov chain with detailed balance |}
The infimum in Eq. \ref{I4ctMC} can be solved if $\{q_{ij}\}$ has detailed balance, ${\pi_iq_{ij}=\pi_jq_{ji}}$:
\begin{equation}
	\sum_{i,j=1}^n\frac{ \mu_j q_{ji}\nu_i }{ \mu_i }  \ge
 \sum_{i=1}^n q_{ii}\nu_i + 2\sum_{i>j}^n\big( q_{ji}\nu_iq_{ij}\nu_j \big)^{\frac{1}{2}},
\end{equation}
where the equality holds true when $\mu_jq_{ji}\nu_i /\mu_i = \mu_iq_{ij}\nu_j/\mu_j$.  That is if $(\mu_j/\mu_i)^2$ $=\nu_j\pi_j/(\nu_i\pi_i) \; \forall i,j$.  Therefore 
\begin{equation}
\frac{ I(\{\nu_i\}) }{\Delta t}
= \sum_{i,j=1}^n \sqrt{\nu_i} \ \hat{\Pi}_{ij}
                \sqrt{\nu_j},
\label{qd}
\end{equation}
in which matrix $\hat{\mPi}$ is semi-positive definite with elements $\hat{\Pi}_{ij}= -\sqrt{\pi}_iq_{ij}/\sqrt{\pi_j}=\hat{\Pi}_{ji}$.  As a LDRF, $I(\{\nu_i\})\ge 0$ and the equality holds true when $\nu_i=\pi_i$.  This corresponds to matrix $\{q_{ij}\}$ having an eigenvalue $0$ with eigenvector $(1,\cdots,1)$.
The Legendre-Fenchel transform of (\ref{qd}) yields an eigenvalue problem that solves the $\{\nu_i\}$ for given $u_i=\partial I/\partial \nu_i$, $(1\le i\le n)$:
\begin{equation}
\label{qd2}
   \sum_{k=1}^n \big[\ \hat{\Pi}_{ik} - (u_i + \lambda )\delta_{ik}\big]
         \sqrt{\nu_k} = 0,
\end{equation}
in which $\lambda$ again represents the gauge freedom that arises from  the constraint among $\nu$'s.  The $\lambda$ is the largest eigenvalue
of matrix $\big[\hat{\mPi}-\text{diag}(u_1,\cdots,u_n)\big]$,
with the corresponding eigenvector being $\{\sqrt{\nu_i}\}$.

%For $2\times 2$ symmetric matrix $\hat{\mPi}$ and homogeneous degree $1$
%\begin{eqnarray*}
%       I(\nu_1,\nu_2) &=&  \hat{\Pi}_{11} \nu_1
%      + 2\sqrt{\nu_1}\ \hat{\Pi}_{12}\sqrt{\nu_2}+
%      \hat{\Pi}_{22} \nu_2,
%\\
%		u_1 &=& \frac{\partial I}{\partial\nu_1 } = \Big(
%             \sqrt{\nu_1}\ \hat{\Pi}_{11}  + 
%              \sqrt{\nu_2}\ \hat{\Pi}_{12}  \Big)\nu_1^{-\frac{1}{2}} ,
%\\
%		\sqrt{\nu_1 } &=&    \hat{\Pi}_{12}\sqrt{\nu_2}
%              ( u_1-\hat{\Pi}_{11} )^{-1},
%\\
%   \tilde{I}(u_1,\nu_2) &=& \left( \hat{\Pi}_{22} -\frac{\hat{\Pi}^2_{12} }{
%              u_1-\hat{\Pi}_{11}}  +  \frac{2\hat{\Pi}^2_{12} }{
%              u_1-\hat{\Pi}_{11}} \right)\nu_2,    
%\end{eqnarray*}
%thus the EoS is $(\hat{\Pi}_{11}-u_1)(\hat{\Pi}_{22} -u_2 ) -\hat{\Pi}_{12}^2= 0$.

Eq. \ref{qd} offers a novel narrative for the statistical behavior of correlated empirical counting frequencies from a time-reversible Markov process: The empirical relative frequencies are split into $\sqrt{\nu_i}$ and $\sqrt{\nu_j}$, they are captured by a symmetric matrix $\sqrt{\nu_i\nu_j}$ whose trace is $1$. This resembles von Neumann's density operator. (\ref{qd2}) further suggests a novel, symmetric eigenvalue problem for $\sqrt{\nu_i}$.  The physical significance of this set of results remains to be explored.

{\bf\em The nature of the Gibbsian theory |} 
We identify Gibbs' theories of statistical mechanics and chemical thermodynamics, as a whole, as a very different program from Boltzmann's original objective of developing a mechanical theory of heat, e.g., thermodynamic principles as mathematical consequences of classical mechanics. Gibbs' theory, with certain justifications, puts the notion of temperature which represents stochastic mechanical motions as an {\it a priori} concept and canonical ensemble theory as a first principle.  This theory reproduces all the known relationships in classical equilibrium thermodynamics by taking the {\em thermodynamic limit}; one of the key results in this regard is the Gibbs-Duhem equation.  However, the new theory itself is applicable to ``small systems''.  The entirety of statistical chemistry and soft matter physics, including theories of non-ideal solutions, polymers, polyelectrolytes, proteins, and the great successes these applications have brought about, is testimony to the validity of the theory, which is far beyond Boltzmann's original vision, if one can surmise. {\em Boltzmann's theory has a goal; Gibbs' is open ended.}

The macroscopic limit gives rise to theories of emergent phenomena, such as phase transitions.  This insight originated in Yang and Lee's work in 1952 \cite{yanglee}, who departed from then traditional approach of McMillan-Mayer cluster expansion.  By 1972, this had become one of the fundamental tenets for emergence \cite{pwanderson,vulpiani-book}.  

The current work finds the Gibbsian theory a new foundation and its limitation. It is a mathematical consequence within the theory of probability, not due to the large size thermodynamic limit, but rather the limit law of large numbers of measurements whose empirical frequency converges to probability.  In other words, the ``law'' articulated in the Gibbsian theory is due to universal statistical behavior of frequency as an empirical measurement for probability, a mathematical rather than physical notion.  

{\bf\em Maximum entropy principle re-interpreted |}
Maximum entropy principle (MEP) \cite{jaynes} and its dynamic counterpart maximum calibre (MaxCal) \cite{dill} have a great number of followers in engineering and data science.  The theory of large deviations offers the very statistical origin for such a principle: The key is to distinguish the abstract mathematical concept of ``probability meansure'' associated with $(\Omega,\mathcal{F},\mathbb{P})$ and the statistical counting frequency that is widely used as the surrogate
for the former.  The latter can be experimentally measured. Very recently, K. A. Dill has called for a distinction between entropy as a scalar quantity of a system and entropy as a ``function of the relative frequencies''.\footnote{https://math.ucr.edu/home/baez/SMB2021/SMB2021\_dill.pdf}

{\bf\em Gauge freedom, normalization and free energy of ground state.}  The Legendre-Fenchel dual to the Shannon-Sanov entropy of counting frequency $\{\nu_i\}$ naturally gives rise to the notion of Gibbs-Kirkwood potential function $\{u_i\}$ together with a gauge freedom $\lambda$ which is mathematically related to the normalization of the $\nu$'s. It defines through a gauge fixing the ``ground'' state in which its Helmholtz free energy $\lambda$ is zero.  It is important to recognize that the set of $\{\nu_i\}$ is a result of measurements approaching infinity in number: Given a set of $\{\nu_i\}$ or alternatively given a set of $\{u_i\}$, one defines a statistical or ``thermodynamic'' system with permanence. In classical physics, such a system is defined mechanically through its energy function.  In the present work, it is replaced by the underlying probability and the LD entropy function.

{\bf\em Bottom up, top down, and bottom down |}  
Gibbs' theory, as understood from the present work, provides a surprisingly penetrating analytic device for carrying out the vision of R. B. Laughlin \cite{laughlin}:  In the past, the theory at the ``bottom'' has always been {\em mechanics}, which provides the notion of mechanical energy and mechanical work, from which thermodynamics arises.  But as many practitioners of biophysical chemistry have recognized \cite{schellman}, the relationship between the macromolecular Boltzmann weights (or Kirkwood's potential of mean force) one has at hand and the constitutive mechanical Hamiltonian of point masses is strenuous even at its best.  The present work, in agreement with \cite{esmith}, again suggests the traditional view might not be the necessary path to thermodynamic behavior.  Statistical law could provide an alternative origin for the very notion of {\em internal energy}, via the Gibbs-Kirkwood potential function.  Further investigations into the counting frequency of correlated events might reveal more unexpected behavior \cite{hoffmann}.

{\bf\em Acknowledgement |}  We thank R. H. Austin, J. C. Baez, K. A. Dill, C. Jia, and H. Zhao for helpful discussions.

\end{document}